\documentclass[aps,prc,twocolumn,superscriptaddress,nofootinbib,showpacs,showkeys,preprintnumbers]{revtex4-1}
\usepackage{graphicx}  % needed for figures
\usepackage{bm}        % for math
\usepackage{amssymb}   % for math
\usepackage{amsmath}   % for math
\usepackage{hyperref}
\usepackage{lineno}

\usepackage[usenames,dvipsnames]{xcolor}

\def\snn{\mbox{$\sqrt{s_{_{\rm NN}}}$}}
\def\AFTER{\mbox{AFTER@LHC}}
\def\etaref{\mbox{$\eta_{\mathrm{ref}}$}}
\def\etasref{\mbox{$\eta_{s,\mathrm{ref}}$}}
\def\trento{T$_{\rm R}$ENTo}

\begin{document}
\title{Anisotropic flow decorrelation in heavy-ion collisions with event-by-event viscous hydrodynamics}
\author{Jakub Cimerman} \affiliation{Faculty of Nuclear Sciences and Physical Engineering, Czech Technical University in Prague,\\  B\v rehov\'a 7, 11519 Prague 1, Czech Republic}
\author{Iurii Karpenko} \affiliation{Faculty of Nuclear Sciences and Physical Engineering, Czech Technical University in Prague,\\  B\v rehov\'a 7, 11519 Prague 1, Czech Republic}
\author{Boris Tom\'a\v{s}ik} \affiliation{Faculty of Nuclear Sciences and Physical Engineering, Czech Technical University in Prague,\\  B\v rehov\'a 7, 11519 Prague 1, Czech Republic}
\affiliation{Univerzita Mateja Bela, Tajovsk\'eho 40, 974~01 Banská Bystrica, Slovakia}
\author{Barbara Antonina Trzeciak} \affiliation{Faculty of Nuclear Sciences and Physical Engineering, Czech Technical University in Prague,\\  B\v rehov\'a 7, 11519 Prague 1, Czech Republic}

\begin{abstract}
Decorrelation of the elliptic flow in rapidity is calculated within a hybrid
approach which includes event-by-event viscous fluid dynamics and final 
state hadronic cascade model. The simulations are performed for Au+Au
collisions at center-of-mass collision energies of 27 and 200 GeV per 
nucleon pair, as well as various asymmetric colliding systems at 72 GeV per nucleon 
pair. Initial conditions determined by an extended Monte Carlo Glauber 
model show better agreement with experimental data than initial conditions
from the UrQMD transport model. We show how the observed decorrelation is 
connected with the decorrelation of initial state spatial anisotropies.
We also study how the effect is increased by the final state hadronic 
cascade. 
\end{abstract}
\maketitle

\section{Introduction}

One of the main objectives of nuclear physics in the last decades has been  the research of deconfined strongly interacting matter called Quark-Gluon Plasma (QGP), which filled the whole Universe microseconds after the Big Bang. This new state of matter has been successfully created experimentally in ultra-relativistic heavy-ion collisions, which allow us to study conditions similar to those at the beginning of the time. 

Especially interesting is the hunt for the critical endpoint, which separates the first order and crossover phase transition between QGP and hadronic matter. Experimentally, this is realised by studying  heavy-ion collisions at energies from a few to tens of GeV, instead of LHC or top RHIC energies. This is the objective of the RHIC Beam Energy Scan (BES) program, the NA62 experiment at CERN, and two facilities under construction: JINR NICA and GSI FAIR.

Hydrodynamics is used to describe the evolution of QGP in heavy-ion collisions since Landau and Bjorken \cite{Landau:1953gs, Bjorken:1982qr}. However, the pure hydrodynamic models are rarely used to describe experimental data nowadays. Instead, hybrid models are broadly used, because they better describe the hadron phase and freeze-out while using event-by-event initial conditions.

Investigations of anisotropic flows $v_n$ can provide us necessary information about the initial state and the evolution of the heavy-ion collision. Most studies focus on the flow in the transverse plane at $y=0$. However studying event-by-event fluctuations along the longitudinal direction is interesting as well. There are arguments that the longitudinal structure of the flow may help us understand the transport properties of QGP \cite{Wu:2018cpc}. 

Flow decorrelation has been studied since a few years, now. Early studies focused on the fluctuations of anisotropic flow along the longitudinal direction \cite{Petersen:2011fp, Pang:2012he}. A linear twist of the event-plane angle $\Psi_n(\eta)$ in the longitudinal direction has been suggested in CGC model \cite{Adil:2005bb, Adil:2005qn} and in Monte-Carlo implementation of the wounded nucleon model \cite{Bozek:2010vz}. The first experimental measurement of the flow decorrelation has been performed by the CMS collaboration \cite{Khachatryan:2015oea} for Pb-Pb collisions at $\snn=2.76$~TeV and $p$-Pb collisions at $\snn=5.02$~TeV. Later, ATLAS collaboration published experimental data at the same energy \cite{Aaboud:2017tql} and for Xe-Xe collisions at $\snn=5.44$~TeV \cite{Aad:2020gfz}. At the time, STAR collaboration shared only preliminary Au+Au results at energies $\snn=27$ and 200~GeV \cite{Nie:2019bgd, Nie:2020trj}. Longitudinal flow correlations are studied also by several theory groups. Event-by-event (3+1)D viscous hydrodynamic model CLVisc with AMPT model for initial conditions has been used in \cite{Pang:2014pxa, Pang:2015zrq, Pang:2018zzo, Wu:2018cpc}. Another (3+1)D viscous hydrodynamic model equipped with Monte Carlo Glauber initial conditions has been exploited in \cite{Bozek:2015bna, Chatterjee:2017mhc, Bozek:2017qir}. In \cite{Sakai:2020pjw}, decorrelation was also studied with the help of a  (3+1)D hydrodynamic model. There are also several studies using pure AMPT model \cite{Jia:2014vja, He:2020xps, Xu:2020koy}. Nevertheless, the longitudinal decorrelation of the anisotropic flow has not yet been theoretically investigated for energies lower than $\snn = 200$~GeV.

Hydrodynamic modelling at RHIC BES energies poses challenges, such as inclusion of finite baryon density in the initial state, in fluid dynamical evolution, and for the fluid-to-hadron transition. We use a three-dimensional event-by-event viscous hydrodynamic model vHLLE+UrQMD~\cite{Karpenko:2013wva} as a base, which allows us to simulate heavy-ion collisions at RHIC BES energies. Thanks to that, in this work, we can focus on two collision energies from the RHIC BES range, namely \snn = 27 and 200 GeV, since these are the energies at which the preliminary experimental data of longitudinal decorrelation are available \cite{Nie:2019bgd, Nie:2020trj}. We examine the reproduction of both longitudinal dependence and the longitudinal decorrelation of the anisotropic flow 
%\del{using three-dimensional event-by-event viscous hydrodynamic model~\cite{Karpenko:2013wva}} 
with two different initial state models: UrQMD \cite{Bass:1998ca} and 3D Monte Carlo Glauber (implemented via GLISSANDO 2 code \cite{Rybczynski:2013yba}). 
Both initial states have non-trivial rapidity dependence and include finite baryon and electric charge densities.
Additionally, we 
%\del{use this model to} 
calculate predictions of longitudinal structure observables for the \AFTER\ programme, which is being proposed as a future fixed-target experiment at the LHC~\cite{Brodsky:2012vg,Massacrier:2015qba,Hadjidakis:2018ifr}. We simulate collisions of Pb+Ti, Pb+W and Pb+C at \snn = 72 GeV, which are three setups being considered for the experiment.

%%%%%%%%%%%%%%%%%%%%%%%%%%%%%%%%%%%%%%%%%%%%%%%%%%%%%

\section{The model}

For this study we use the event-by-event viscous hydrodynamic model as in \cite{Cimerman:2020iny}. The model has three stages. For the initial state we used two different models: UrQMD \cite{Bass:1998ca} and GLISSANDO 2 \cite{Rybczynski:2013yba}. In our previous work we also utilized  the \trento\ model, however 
implementing initial state tilt with out-of-the-box \trento\ code is not a straightforward task, 
%\del{it lacks any tilt in the longitudinal direction and therefore the flow would be fully correlated,} 
so we leave this initial state out\footnote{However, few different ways to extend \trento\ IS into the longitudinal direction are presented in \cite{Ke:2016jrd}, where the $r_n$ has been computed among other observables for pPb and PbPb collisions at the LHC energies.}. The hot and dense phase of the evolution is simulated with three-dimensional viscous code vHLLE \cite{Karpenko:2013wva}. From the freeze-out hypersurface, a Cooper-Frye formula \cite{Cooper:1974mv} with corrections due to the shear viscosity is used to sample hadrons. After the particlization, the UrQMD cascade simulates the hadronic rescatterings and resonance decays. The hybrid model is described in more detail in \cite{Karpenko:2015xea}.

%%%%%%%%%%%%%%%%%%

\subsection{Initial states}

\subsubsection{UrQMD}

The first initial state is taken from UrQMD \cite{Bass:1998ca}, which is a microscopic transport model. UrQMD uses PYTHIA6 to simulate initial inelastic nucleon-nucleon scatterings through string formation and subsequent string break-up. This leads to formation of hadrons, which can then rescatter. These rescatterings are allowed until hydrodynamization, which happens at a hypersurface $\tau=\tau_0$, where $\tau=\sqrt{t^2-z^2}$ is longitudinal proper time. The parameter $\tau_0$ depends on collision energy and its value is set  from the  adjustment to the data. 
%\del{We took the values from \cite{Karpenko:2015xea} and they are listed in Table~\ref{tb:long-params}.} 
Each hadron that crosses the hypersurface $\tau=\tau_0$ smoothly deposits its energy and momentum into few neighbouring cells of the hydrodynamic grid with a weight given by Gaussian distribution
\begin{equation}
w \propto \exp\left(-\frac{(x_h-x_c)^2}{R_T^2}-\frac{(y_h-y_c)^2}{R_T^2} - \gamma^2\tau_0^2\frac{(\eta_h-\eta_c)^2}{R_\eta^2}\right),
\end{equation}
where the coordinates with index $h$ are those of the hadron, coordinates with index $c$ are those of the hydrodynamic cell, $\eta_{h/c}$ is a space-time rapidity and $R_T$, $R_\eta$ are the parameters that control how distant cells this hadron can affect in given direction. In this point, the transformation from Cartesian to Milne coordinates is required, since UrQMD uses the former  and vHLLE the latter. During all these processes, the energy, momentum, baryon number and electric charge are conserved. The \snn-dependent values of $\tau_0$, $R_T$ and $R_\eta$ are taken from \cite{Karpenko:2015xea} and the $\tau_0$ values are listed in Table~\ref{tb:long-params}.

\subsubsection{GLISSANDO}

A much simpler initial state model, GLISSANDO 2 \cite{Rybczynski:2013yba} is an implementation of the  Monte Carlo  Glauber model. It generates positions of nucleons in the transverse plane and determines the number of binary collisions suffered by each nucleon. All these create sources of energy depositions.  Following \cite{Bozek:2012fw, Bozek:2015bha}, the entropy density is distributed as
\begin{multline}\label{eq:s-xy-eta}
s(x,y,\eta_s)= \kappa \sum_i f_\pm(\eta_s)\left[ (1-\alpha) + N^{\rm coll}_i \alpha \right]  \\
{} \times \exp\left( - \frac{(x-x_i)^2+(y-y_i)^2}{2\sigma^2}\right),
\end{multline}
where $x,y$ are the coordinates in transverse plane, $\eta_s=1/2\ln((t+z)/(t-z))$ is the space-time rapidity, $\kappa$ is the normalization constant, the sum goes through the participant nucleons, $N^{\rm coll}_i$ is the number of collisions of the participant $i$ and $\sigma=0.4$ is the width of Gaussian smearing. The mixing parameter $\alpha$ regulates the contribution from the participant nucleons and the binary scatterings. We took $\alpha=0.123$ for $\snn=27$~GeV and $\alpha=0.145$ for $\snn=200$~GeV. Since Glauber model is only two-dimensional, it was necessary to add a longitudinal structure. Again, following \cite{Bozek:2012fw, Bozek:2015bha}, we create an approximately triangular shape of the entropy deposition from forward ($+$) and backward-going ($-$) participant nucleons as
\begin{align}
f_{\pm}(\eta_s)&=\frac{\eta_{\rm M}\pm \eta_s}{2 \eta_{\rm M}} H(\eta_s)\ & \mbox {for } \ |\eta_s|<\eta_{\rm M}
\label{eq:lprof}
\end{align}
where $\eta_{\rm M}$ is a parameter of the Bialas-Czyz-Bozek model taken from \cite{Rybczynski:2013yba} determining the longitudinal extent of an entropy density deposition from each participant nucleon, and the profile function $H(\eta_s)$ is defined via
\begin{equation}
H(\eta_s)=\exp\left(-\frac{(|\eta_s|-\eta_0)^2\Theta(|\eta_s|-\eta_0)}{2\sigma_\eta^2}\right) .  \label{eq:Heta}
\end{equation}
Here, $\eta_0$ is the half-width of the plateau in longitudinal direction.

Originally, this hybrid model was developed for higher energies, where the baryon density could be neglected. However, for energies lower than the top RHIC energy, it plays an important role. Thus we added to the model (for energies $\snn=27$ and 72~GeV) also baryon charge deposition from participants
\begin{eqnarray}
n_B(x,y,\eta_s)&=& \kappa_B \sum_i \exp\left( -\frac{(\eta_B\pm\eta_s)^2}{2\sigma_B^2} \right) \times \nonumber \\
&& \exp\left( - \frac{(x-x_i)^2+(y-y_i)^2}{2\sigma^2}\right) . \label{eq:nB-eta}
\end{eqnarray}
This ansatz assumes that the forward-going participants deposit their baryon charge around $+\eta_B$, while backward-going participants around $-\eta_B$. The \snn-dependent values of the longitudinal structure parameters are taken from our previous study \cite{Cimerman:2020iny} and are listed in Table~\ref{tb:long-params}. Note that, at the lowest collision energy of interest \snn=27 GeV, the optimal description of basic experimental observables necessitated centrality-dependent $\eta_0$, $\sigma_\eta$, $\eta_B$ and $\sigma_B$. In practice the centrality dependence is introduced via the centrality measure $\chi=N_W/(2A)$. We have interpreted such centrality dependence in \cite{Cimerman:2020iny} as a change in the strength of baryon stopping with changing nuclear thickness at different collision centralities.

Along with the deposited local baryon density, we need to add also deposition of local electric density, which is given by
\[ n_Q = 0.4\ n_B, \]
where the factor 0.4 is coming from the approximate charge-to-mass ratio in heavy nuclei.
Normalization constants $\kappa$ and $\kappa_B$ were chosen numerically, so that the total energy and the total baryon charge are conserved during the hydrodynamization, i.e.\ after initialisation of the fluid we have
\begin{eqnarray}
\tau_0 \int \epsilon\cosh\eta\, dxdyd\eta & = & \frac{N_\text{W}}{2}\snn \\
\tau_0 \int n_B dxdyd\eta & =& N_\text{W}.
\end{eqnarray}

The values of the parameters  for $\snn=27$~GeV were taken from \cite{Cimerman:2020iny}. Since we did not have any experimental data at $\snn=72$~GeV available, we used the values for $\snn=62.4$~GeV, also for this energy. For $\snn=200$~GeV we took the parameters from \cite{Bozek:2015bha}. Values of parameters for all three energies are summarized in Table~\ref{tb:long-params}.

%%%%%%%%%%%%%%%%%
\begin{table}
\begin{center}
\begin{tabular}{c||c|c|c}
 $\snn$~[GeV] & 27 & 72 & 200 \\ \hline \hline
 $\tau_0$~[fm/c] & 1.0 & 0.7 & 0.4 \\ \hline
 $\eta/s$ & 0.12 & 0.08 & 0.08 \\ \hline
 $\eta_0$ & $0.89-0.2\chi$ & 1.8 & 1.5 \\ \hline
 $\sigma_\eta$ & $1.09-0.2\chi$ & 0.7 & 1.4 \\ \hline
 $\eta_M$ & 1.0 & 1.8 & 3.36 \\ \hline
 $\eta_B$ & $1.33 - 0.32\chi$ & 2.2 \\ \hline
 $\sigma_B$ & $0.79 - 0.21\chi$ & 1.0 \\ 
 \end{tabular}
\caption{Default values of the GLISSANDO model parameters: starting time, shear viscosity and the parameters for the longitudinal profile for $\snn=27$, 72 and 200~GeV. In the $\snn=27$~column,  centrality measure $\chi=N_\text{W}/(2A)$ is introduced.}\label{tb:long-params}
\end{center}
\end{table}
%%%%%%%%%%%%%%%%%%%%%%%

%%%%%%%%%%%%%%%%%%%%%%%%

\subsection{Hydrodynamics}

The hydrodynamic phase of the heavy-ion collision is simulated using three-dimensional relativistic viscous hydrodynamic code vHLLE \cite{Karpenko:2013wva}. This code solves the energy-momentum and baryon number conservation equations
\begin{equation} 
\nabla_{\nu} T^{\mu\nu}=0\nonumber,\quad \nabla_{\nu}n_B^\nu=0.
\end{equation}
The viscous corrections are added within M\"uller-Israel-Stewart framework using the evolution equations for the shear stress tensor
\begin{equation}
\left\langle u^\gamma \nabla_{\gamma} \pi^{\mu\nu}\right\rangle =-\frac{\pi^{\mu\nu}-\pi_\text{NS}^{\mu\nu}}{\tau_\pi}-\frac 4 3 \pi^{\mu\nu}\nabla_{\gamma}u^\gamma
\end{equation}
where $\pi_\text{NS}^{\mu\nu}$ is the shear stress tensor from the Navier-Stokes limit. For this study, we used the values of the shear viscosity from \cite{Karpenko:2015xea} (see Table \ref{tb:long-params}), while we set zero bulk viscosity.

Our hydrodynamic model uses a chiral model for the Equation of State (EoS) \cite{Steinheimer:2010ib}. This EoS has a crossover phase transition between hadron matter and QGP. Its results qualitatively agree with the lattice QCD calculations at $\mu_B=0$.

%%%%%%%%%%%%%%%%%%%%%%%%%%%%%%%%%%%%%%%

\subsection{Particlization and freeze-out}

The fluid-to-particle transition happens at the hypersurface with fixed energy density $\epsilon=0.5$~GeV/fm$^3$, which is found using the Cornelius subroutine \cite{Huovinen:2012is}. The number of particles, which are emitted on the freeze-out hypersurface $\Sigma$, is given by the Cooper-Frye formula \cite{Cooper:1974mv}
\begin{equation}
N=\int\frac{d^3p}{E_p}\int d\Sigma_\mu(x)p^\mu f(x,p),
\end{equation}
where $f(x,p)$ is the phase-space distribution function of non-interacting hadrons and hadronic resonances. This formula is then supplemented with the ansatz for viscous corrections. To better reproduce the experimental setup, the model samples the produced particles using a Monte Carlo procedure instead of calculating the Cooper-Frye integrals directly. The overall formula for hadron sampling is given by
\begin{multline}
\frac{d^3 \Delta N_i}{dp^* d({\rm cos}\,\theta)d\phi}=\frac{\Delta\sigma^*_\mu p^{*\mu}}{p^{*0}} 
p^{*2} f_\text{eq}(p^{*0};T,\mu_i) \\
\times\left[ 1+(1\mp f_\text{eq})\frac{p^*_\mu p^*_\nu \pi^{*\mu\nu}}{2T^2(\epsilon+p)} \right]. \label{DF-LRF-visc}
\end{multline}
The final step of the model is to simulate hadronic rescatterings and resonance decays of the sampled hadrons using UrQMD cascade \cite{Bass:1998ca}.

To increase the statistics of the final-state hadronic events, we use oversampling of the hadrons coming from each hydrodynamic configuration in the event-by-event ensemble. We sample hadrons using Cooper-Frye formula few hundreds of times from each resulting freeze-out hypersurface, and then pass these events separately to the UrQMD cascade.

%%%%%%%%%%%%%%%%%%%%%%%%%%%%%%%%%%%%%%%%%%%%%%%%%%%%%%%%%%%%%%%%%%%%%%%%

\section{Results}

In this study we focus on one wide centrality interval $10-40\%$, since most of the experimental data is available in this centrality bin. We simulated heavy-ion collisions at two collision energies where the experimental data exist: $\snn=27$ and 200~GeV.

\subsection{Rapidity dependence of elliptic flow}

\begin{figure}[tbh]
    \centering
    \includegraphics[width=0.5\textwidth]{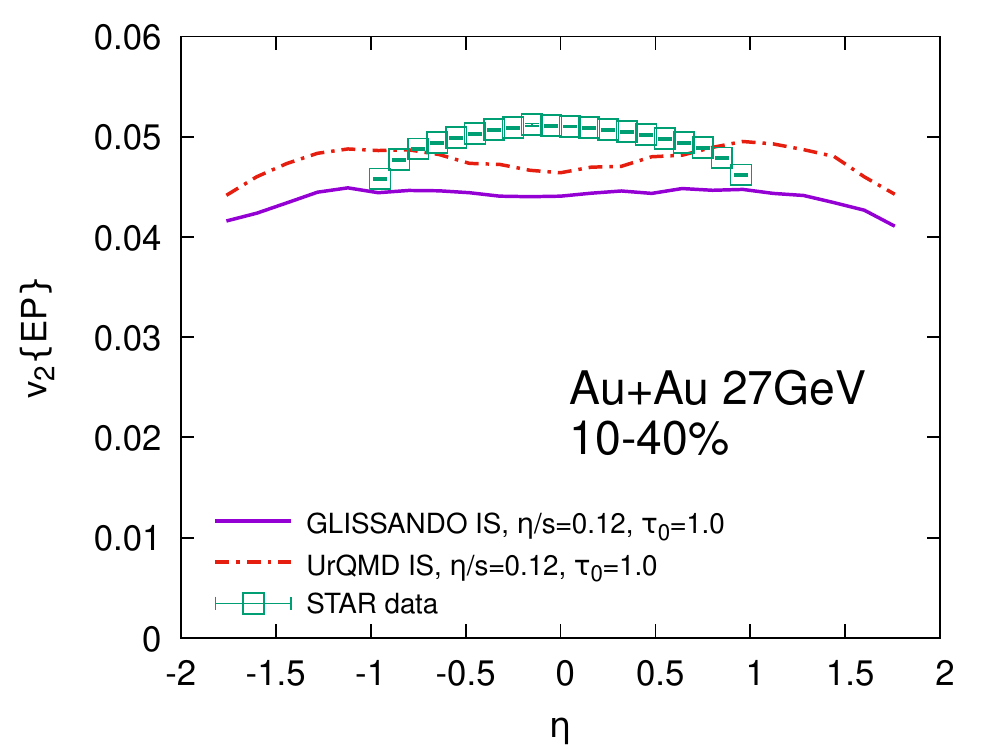}
    \caption{Elliptic flow as a function of pseudorapidity for 10-40\% Au-Au collisions at $\snn=27$~GeV from vHLLE+UrQMD simulations with UrQMD and GLISSANDO initial states. The experimental data points are taken from \cite{Adamczyk:2012ku}.}
    \label{fig:v2-27}
\end{figure}

\begin{figure}[tbh]
    \centering
    \includegraphics[width=0.5\textwidth]{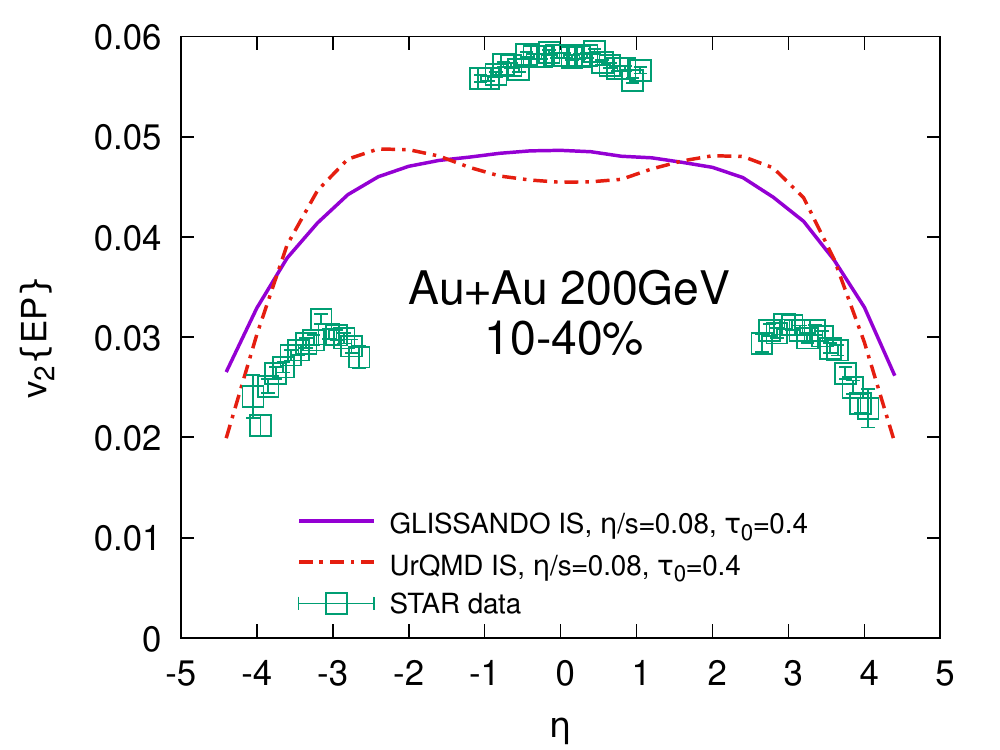}
    \caption{Same as Fig. \ref{fig:v2-27} but for $\snn=200$~GeV. The experimental data points are taken from \cite{Abelev:2008ae}.}
    \label{fig:v2-200}
\end{figure}

First, we show the pseudorapidity dependence of the elliptic flow of charged hadrons. Figures \ref{fig:v2-27} and \ref{fig:v2-200} show the elliptic flow as a function of pseudorapidity for both UrQMD initial conditions and GLISSANDO initial conditions. We use the event-plane (EP) method \cite{Holopainen:2010gz} for this calculation and compare our results to $v_2\{\textrm{EP}\}$ from STAR~\cite{Adamczyk:2012ku,Abelev:2008ae}. At $\snn=27$~GeV both initial states reproduce overall order of magnitude of the elliptic flow, but underestimate its value at mid-rapidity, which is consistent with our previous results \cite{Cimerman:2020iny}. At $\snn=200$~GeV the experimental data indicate triangular pseudo-rapidity dependence that neither of the initial states can describe. A similar triangular shape is reported also in older experimental data from PHOBOS~\cite{Back:2004mh}. This  shape was reproduced by hydrodynamic simulations  in~\cite{Nonaka:2006yn}, however, an EoS with 1$^{\rm st}$-order phase transition was used there, which resulted in slower expansion as compared to our calculation.

%%%%%%%%%%%%%%%%%%%%%%%%%%

\subsection{Longitudinal decorrelation of the anisotropic flow}

For the calculation of the flow decorrelation we first need define the flow vector:
\begin{equation}
q_n(\eta)=\frac{1}{m}\sum_{k=1}^m e^{in\phi_k}=v_n(\eta)e^{in\Psi_n(\eta)},
\label{eq:qn}
\end{equation}
where $m$ is the number or charged hadrons in the examined pseudorapidity interval, $\phi_k$ is the azimuthal angle of hadron momentum, $v_n$ is the magnitude of the flow and $\Psi_n$ is the corresponding event-plane angle.

Using the flow vector we write the factorization ratio $r_n$ as
\begin{equation}
r_n(\eta)=\frac{\left\langle q_n(-\eta)q_n^\ast(\etaref) \right\rangle}{\left\langle q_n(\eta)q_n^\ast(\etaref) \right\rangle}.
\label{eq:rn}
\end{equation}
Here, $\left\langle \dots \right\rangle$ denotes averaging over events and $\etaref$ is the reference bin chosen far enough in forward or backward pseudorapidity region so it does not overlap with the $\eta$ bins. For comparison with STAR preliminary data \cite{Nie:2019bgd, Nie:2020trj}, we used the same pseudorapidity reference bins, namely $2.1<\etaref<5.1$ for $\snn=27$~GeV Au+Au collisions and $2.5<\etaref<4$ for $\snn=200$~GeV Au+Au collisions. We also used the same $p_T$ cut for charged hadrons ($0.4<p_T<4$~GeV/$c$). However, since these are just preliminary data, we were not able to extract error bars from the plots. 

\begin{figure}[tbh]
\centering
\includegraphics[width=0.5\textwidth]{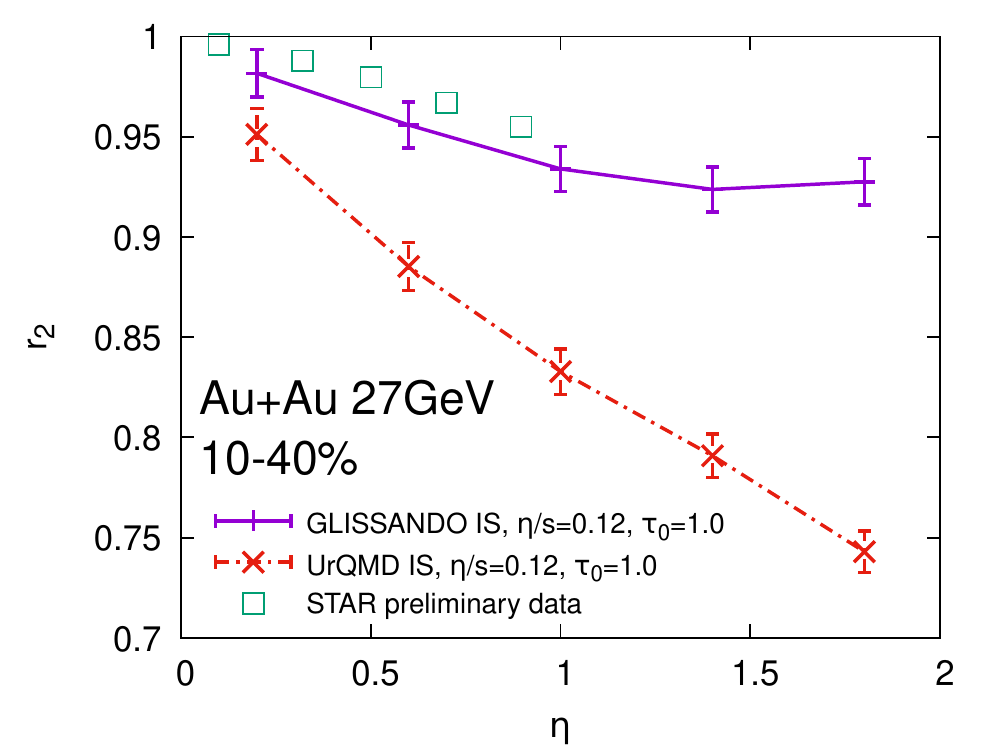}
\caption{The factorization ratio $r_2$ as a function of pseudorapidity for 10-40\% Au-Au collisions at $\snn=27$~GeV from vHLLE+UrQMD simulations with UrQMD and GLISSANDO initial states. The preliminary experimental data points are taken from \cite{Nie:2020trj}.}
\label{fig:r2-27}
\end{figure}

\begin{figure}[tbh]
\centering
\includegraphics[width=0.5\textwidth]{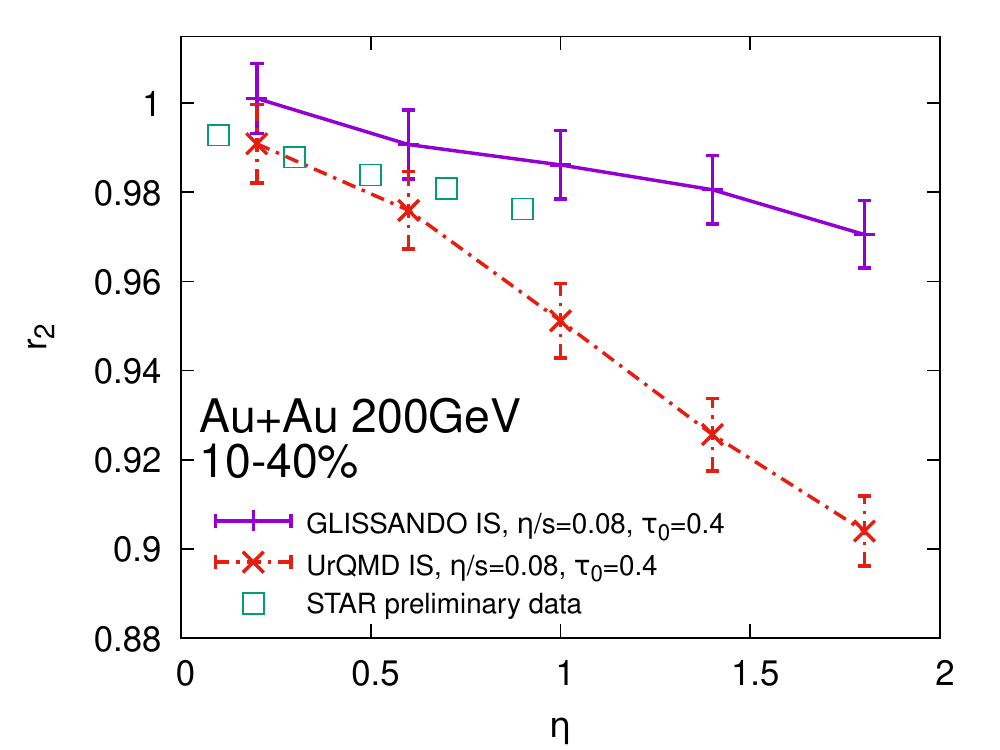}
\caption{Same as Fig. \ref{fig:r2-27} but for $\snn=200$~GeV. The preliminary experimental data points are taken from \cite{Nie:2019bgd}.}
\label{fig:r2-200}
\end{figure}

Figures \ref{fig:r2-27} and \ref{fig:r2-200} show our results of flow decorrelation factorization ratio in comparison with STAR preliminary data. Figure \ref{fig:r2-27} indicates that at $\snn=27$~GeV, UrQMD creates much stronger decorrelation than the one seen in the data. On the other hand, calculations with GLISSANDO can describe the experimental data within uncertainties for both energies. 

\begin{figure}[tbh]
\centering
\includegraphics[width=0.5\textwidth]{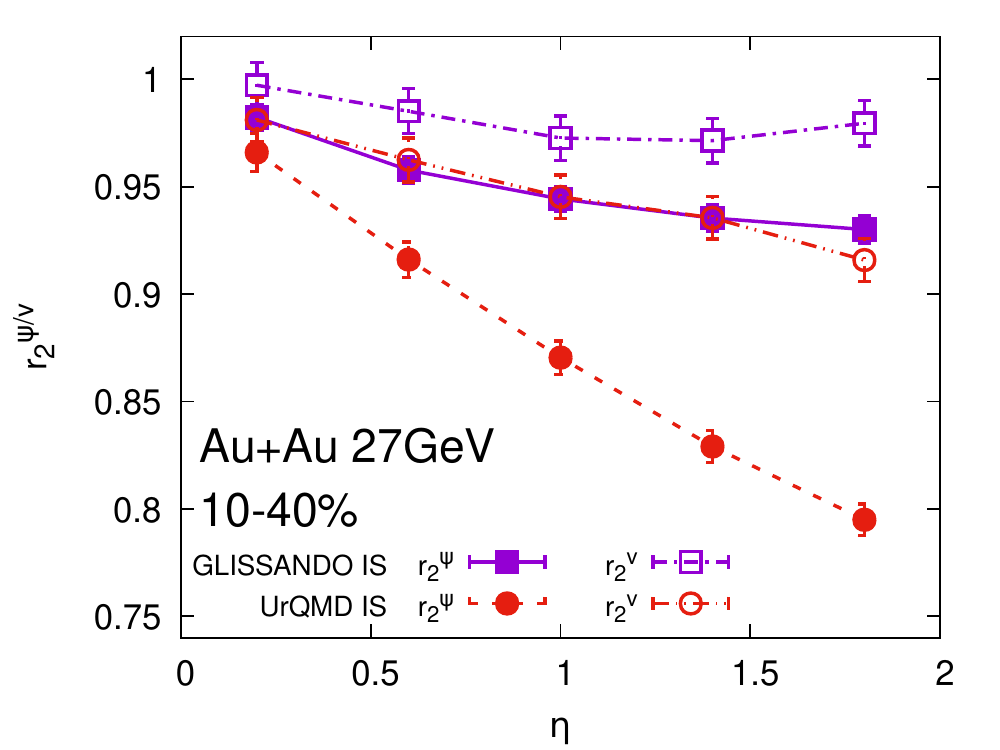}
\caption{The flow magnitude decorrelation $r_2^v$ and the flow angle decorrelation $r_2^\Psi$ as a function of pseudorapidity for 10-40\% Au-Au collisions at $\snn=27$~GeV from vHLLE+UrQMD simulations with UrQMD and GLISSANDO initial states.}
\label{fig:r2-v/psi}
\end{figure}

In order to better understand where the decorrelation is coming from, 
equation \eqref{eq:rn} can be rewritten using the definition of the flow vector as follows:
\begin{equation}
r_n(\eta)=\frac{\left\langle v_n(-\eta)v_n(\etaref) \cos[n\left(\Psi_n(-\eta)-\Psi_n(\etaref)\right)]\right\rangle}{\left\langle v_n(\eta)v_n(\etaref) \cos[n\left(\Psi_n(\eta)-\Psi_n(\etaref)\right)]\right\rangle}.
\label{eq:rn-long}
\end{equation}
From this formula it can be easily seen that the flow decorrelation may be caused by two separate effects: flow magnitude decorrelation and flow angle decorrelation. The former effect alone describes a case, where the flow magnitude is uncorrelated between $\eta$ and $-\eta$, while event-plane angle stays the same. We can calculate it as
\begin{equation}
r_n^v(\eta)=\frac{\left\langle v_n(-\eta)v_n(\etaref)\right\rangle}{\left\langle v_n(\eta)v_n(\etaref) \right\rangle}.
\label{eq:rn-v}
\end{equation}
The latter one describes a case of fully correlated magnitude, but uncorrelated event-plane angle. This can be calculated via
\begin{equation}
r_n^\Psi(\eta)=\frac{\left\langle \cos[n\left(\Psi_n(-\eta)-\Psi_n(\etaref)\right)]\right\rangle}{\left\langle  \cos[n\left(\Psi_n(\eta)-\Psi_n(\etaref)\right)]\right\rangle}.
\label{eq:rn-psi}
\end{equation}
It has already been shown for LHC and top RHIC energies that these two effects do not contribute to the resulting factorization ratio equally---the flow angle decorrelation dominates \cite{Bozek:2017qir, Wu:2018cpc}. Thus, we calculated these two contributions for $\snn=27$~GeV. The result is shown in Fig. \ref{fig:r2-v/psi}. We can confirm, that even at much lower energy, the flow angle decorrelation still plays more important role than the flow magnitude decorrelation, which has been shown by both initial state models. 

%%%%%%%%%%%%%%%%%%%%

\subsection{Locality in the UrQMD initial state}

\begin{figure}[tb]
\centering
\includegraphics[width=0.5\textwidth]{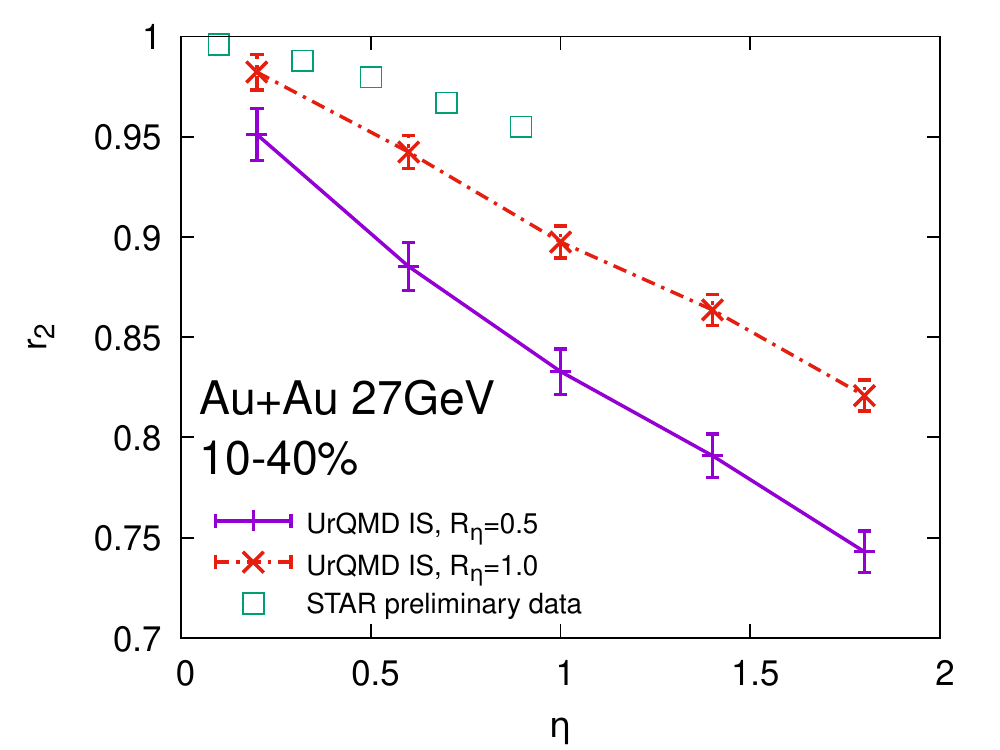}
\caption{The effect of parameter $R_\eta$ on the factorization ratio $r_2$ for 10-40\% Au-Au collisions at $\snn=27$~GeV from vHLLE+UrQMD simulations with UrQMD initial state. The preliminary experimental data points are taken from \cite{Nie:2020trj}.}
\label{fig:r2-rgz}
\end{figure}

\begin{figure}[tb]
\centering
\includegraphics[width=0.5\textwidth]{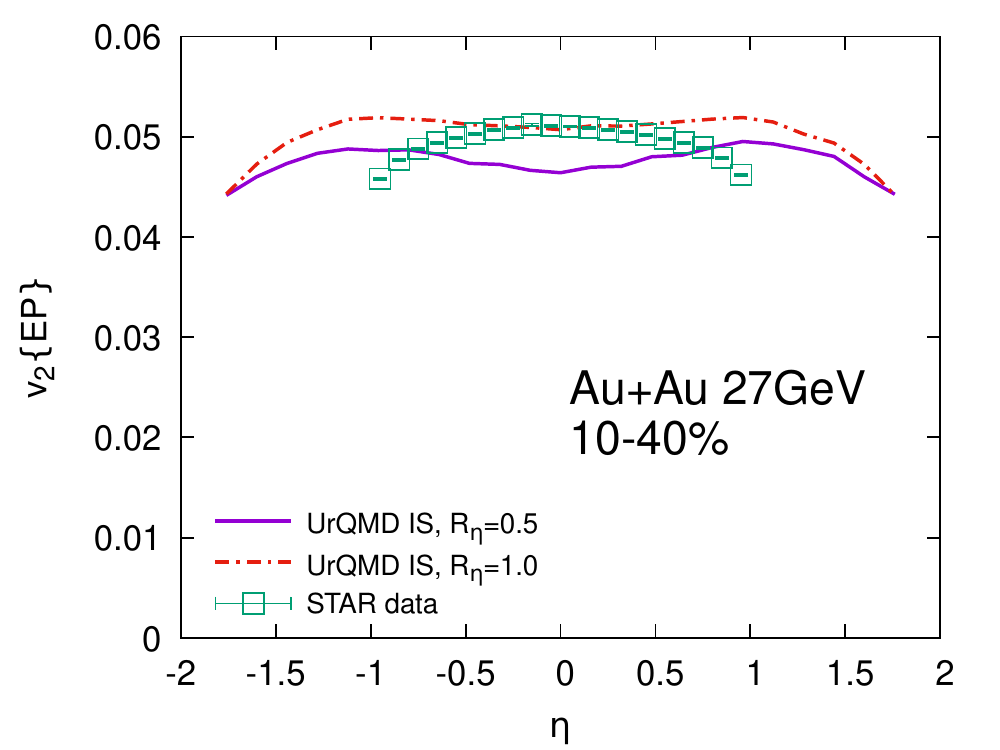}
\caption{Same as Fig. \ref{fig:r2-rgz} but for elliptic flow. The experimental data points are taken from \cite{Adamczyk:2012ku}.}
\label{fig:v2-rgz}
\end{figure}

The results with UrQMD IS show much stronger decorrelation than can be seen in the data. This can be understood so that the UrQMD initial state is too local rapidity-wise. The deposition from each nucleon-nucleon scattering is quite narrow in space-time rapidity as compared to the scaled 3D Glauber model with the triangular rapidity deposition. At this point we attempt to improve the locality of UrQMD IS by increasing the value of parameter $R_\eta$, which controls the smearing of the energy-momentum deposition from each initial state hadron in the space-time rapidity. Figures \ref{fig:r2-rgz} and \ref{fig:v2-rgz} show the comparison of calculations with the default value $R_\eta=0.5$ and increased value $R_\eta=1.0$ at $\snn=27$~GeV in the factorization ratio and the elliptic flow, respectively. We can see that such modification increases the value of the factorization ratio, which then becomes closer to the experimental data. Moreover, it increases the $v_2$ a bit, so now the mid-rapidity value start to agree with the experimental data.

\begin{figure}[tb]
\centering
\includegraphics[width=0.5\textwidth]{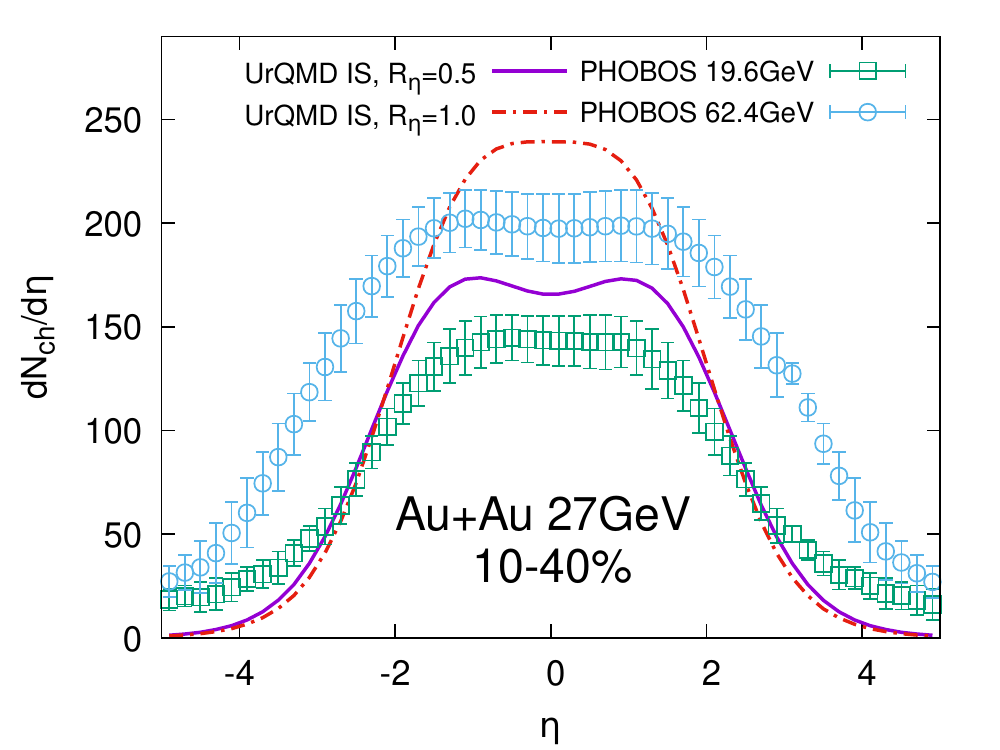}
\caption{Same as Fig. \ref{fig:r2-rgz} but for pseudorapidity distribution. Since there are no experimental data for $\snn=27$~GeV, we compare the distribution to experimental data from PHOBOS at energies $\snn=19.6$ and 62.4~GeV taken from \cite{Alver:2010ck}.}
\label{fig:dndeta-rgz}
\end{figure}

The caveat is that the $R_\eta=1.0$ setting fails to reproduce the more basic observable---the pseudorapidity distribution. Figure \ref{fig:dndeta-rgz} shows the  $dN/d\eta$ with $R_\eta=0.5$ and $R_\eta=1.0$ settings. It is obvious that the result with $R_\eta=1.0$ significantly overestimates the overall multiplicity and thus we cannot use this setting any further.

%%%%%%%%%%%%%%%%%%%

\subsection{Impact of the final-state rescatterings}

\begin{figure}[tb]
\centering
\includegraphics[width=0.5\textwidth]{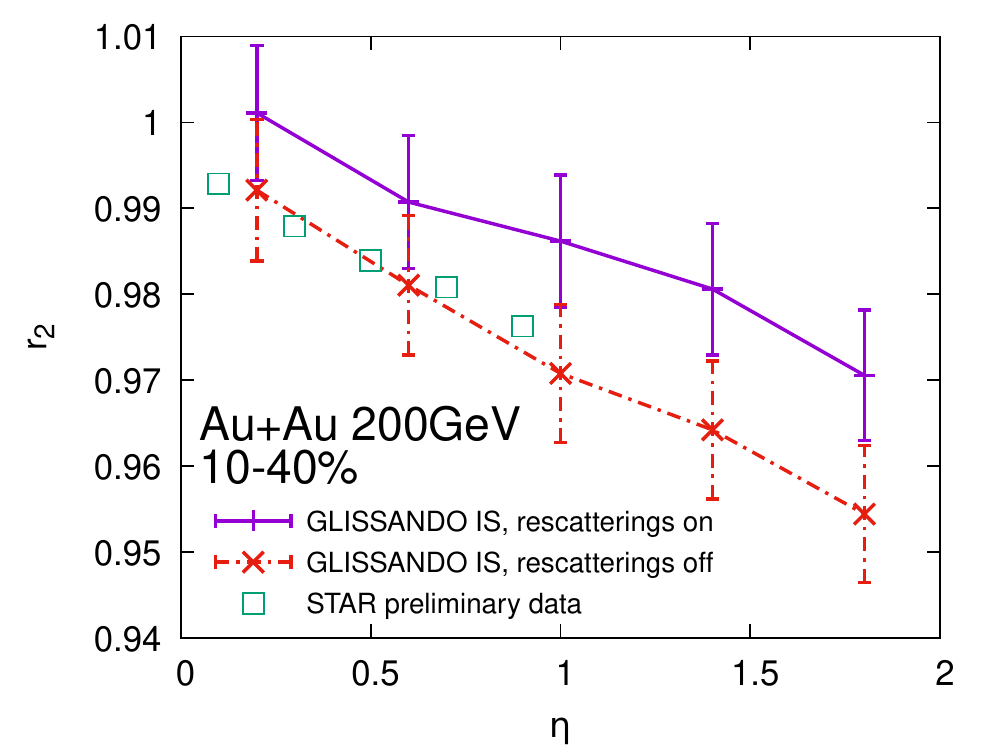}
\caption{The effect of final-state rescatterings on the factorization ratio $r_2$ for 10-40\% Au-Au collisions at $\snn=200$~GeV from vHLLE+UrQMD simulations with UrQMD initial state. The preliminary experimental data points are taken from \cite{Nie:2019bgd}.}
\label{fig:r2-rescat}
\end{figure}

\begin{figure}[tbh]
\centering
\includegraphics[width=0.5\textwidth]{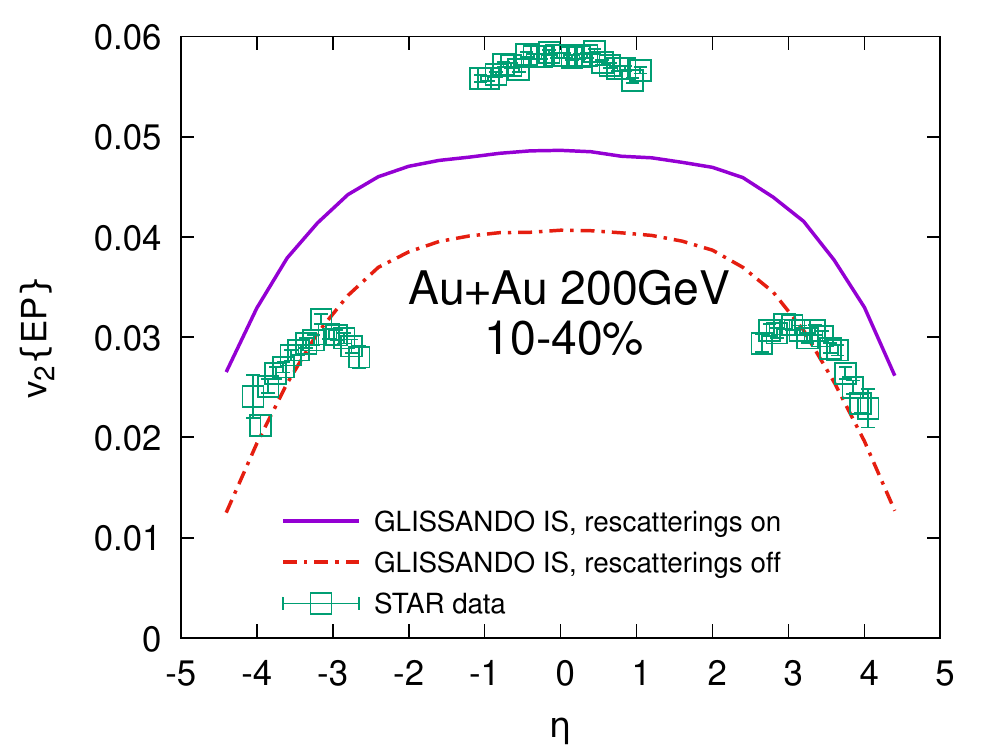}
\caption{Same as Fig. \ref{fig:r2-rescat} but for elliptic flow. The experimental data points are taken from \cite{Abelev:2008ae}.}
\label{fig:v2-rescat}
\end{figure}

Our calculations are performed with the final-state rescatterings simulated by UrQMD cascade \cite{Bass:1998ca}. However, the so-far published hydrodynamic calculations of the flow decorrelation \cite{Bozek:2017qir, Pang:2015zrq, Pang:2018zzo, Wu:2018cpc} were done without hadronic rescatterings in the post-hydro phase. Therefore, it is instructive to see, how that can affect the factorization ratio. Therefore, we run another calculation with final-state hadronic rescatterings via UrQMD turned off, for Au-Au collisions at $\snn=200$~GeV with GLISSANDO initial state. These results are shown in Fig. \ref{fig:r2-rescat} for the flow decorrelation and Fig. \ref{fig:v2-rescat} for the elliptic flow. It can be seen that turning off the rescatterings causes stronger decorrelation, which brings the simulations closer to the experimental data, and it also causes a decrease of the elliptic flow. Note that an increase of the elliptic flow via the final-state hadronic rescatterings is a well-known effect, reported in an early application of a hybrid model to $\snn=200$~GeV RHIC data \cite{Song:2010aq}.

%%%%%%%%%%%%%%%%%%%%

\subsection{Initial state eccentricity}

\begin{figure}[tb]
\centering
\includegraphics[width=0.5\textwidth]{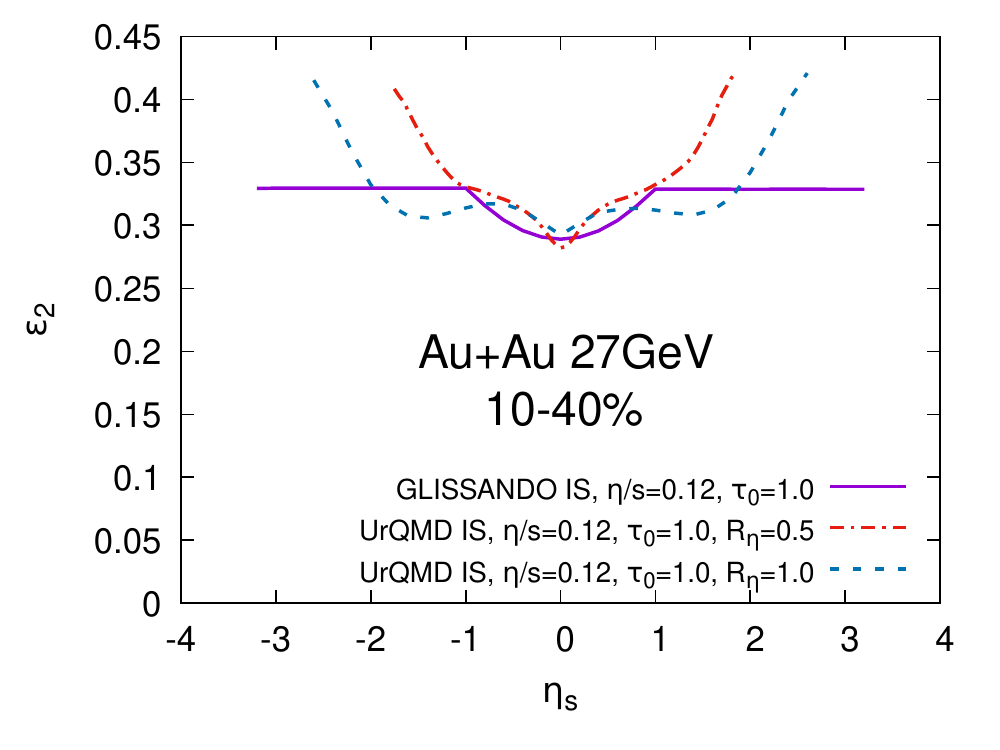}
\caption{Initial state eccentricity $\epsilon_2$ as a function of space-time rapidity for 10-40\% Au-Au collisions at $\snn=27$~GeV with UrQMD and GLISSANDO initial states.}
\label{fig:eps2-27}
\end{figure}

\begin{figure}[tb]
\centering
\includegraphics[width=0.5\textwidth]{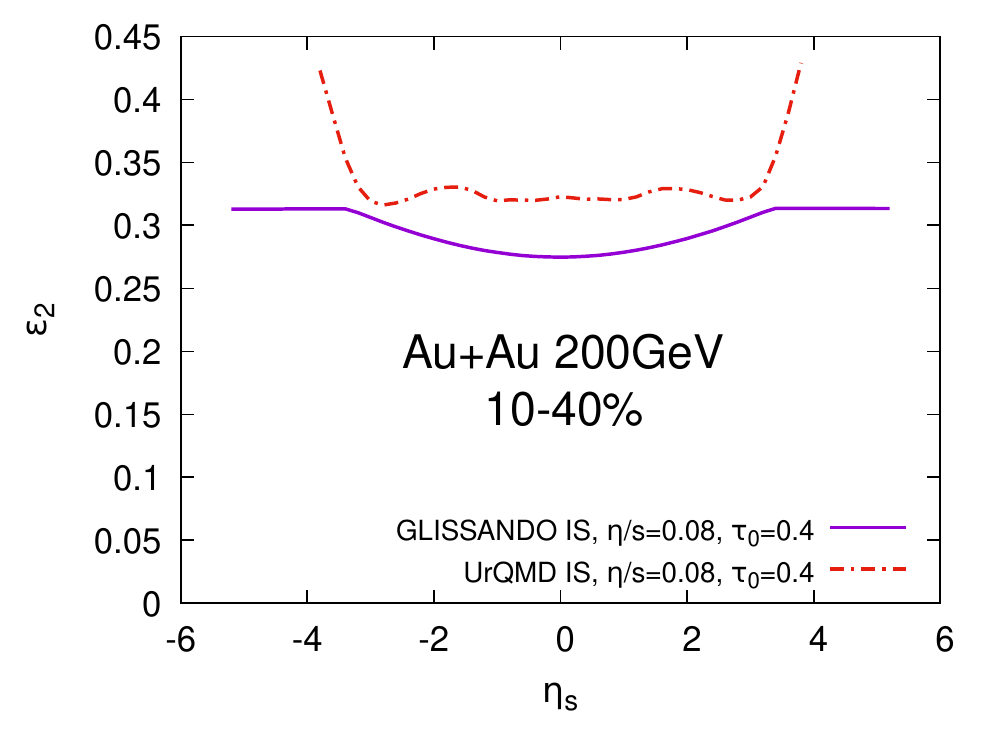}
\caption{Same as Fig. \ref{fig:eps2-27} but for $\snn=200$~GeV.}
\label{fig:eps2-200}
\end{figure}

It is well known that the anisotropy coefficients $v_2$ are strongly 
correlated with the initial state anisotropies of the spatial distribution of the hot matter \cite{Niemi:2012aj}. Therefore, with both initial state models we calculate the initial state eccentricity at the moment when fluid-dynamical simulation is started.
%Next, we looked at the initial state anisotropy. 
The $n$-th order eccentricity of the initial state is defined as
\begin{equation}\label{eq:eps-eta}
\epsilon_n e^{in\Psi_n}=\frac{\int e^{in\phi}r^n\rho(\vec{r}) d\phi\, r \,dr}{\int r^n \rho(\vec{r}) d\phi\, r \,dr},
\end{equation}
where $\vec{r}=(\tilde{x},\tilde{y})$ is the position vector in transverse plane with the correction for the center of mass ($\tilde{x}=x-x_{cm}$, $\tilde{y}=y-y_{cm}$). 
%\del{The pseudorapidity dependence of $\epsilon_2$ is shown in Figs.\ \ref{fig:eps2-27} and \ref{fig:eps2-200}.} 
We apply Eq.~\ref{eq:eps-eta} for different space-time rapidity slices and the resulting space-time rapidity dependence is shown in Figs. \ref{fig:eps2-27} and \ref{fig:eps2-200}. 
%\del{In these plots we can see the initial anisotropy of both initial states and the difference with which they enter to the hydrodynamics.}
One can observe a somewhat different $\eta_s$ dependence of $\epsilon_2$ for the two different initial state models. In UrQMD IS, the hadrons end up at different space-time rapidities at $\tau=\tau_0$ hypersurface due to finite thickness of the incoming nuclei and secondary scatterings. There are fewer hadrons at large $\eta_s$, which leads to larger $\epsilon_2$ and larger associated fluctuations in the $\Psi_n$ angle. Different from that, in the 3D GLISSANDO IS, the transverse entropy density profile at each space-time rapidity is formed either from projectile or target wounded nucleons, or from a mixture of both. The $f_\pm(\eta_s)$ in Eq.~\ref{eq:s-xy-eta} acts as a mixing factor, which retains only projectile and only target nucleons beyond $\pm\eta_M$ respectively. Therefore, for $|\eta_s|>\eta_M$ the transverse shape of the density profile does not change anymore, and as such the $\epsilon_2$ remains constant as one can see in Figs. \ref{fig:eps2-27} and \ref{fig:eps2-200}.

\begin{figure}[tb]
\centering
\includegraphics[width=0.5\textwidth]{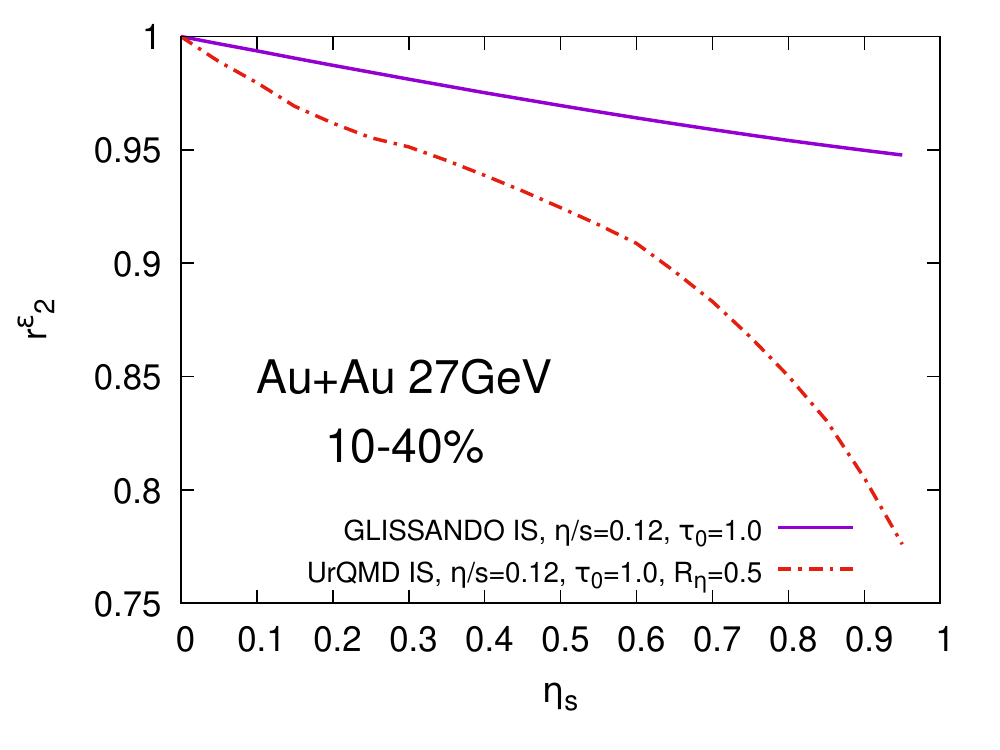}
\caption{Longitudinal decorrelation of the initial state eccentricity $\epsilon_2$ as a function of space-time rapidity for 10-40\% Au-Au collisions at $\snn=27$~GeV with UrQMD and GLISSANDO initial states.}
\label{fig:eps2-decorrealtion}
\end{figure}

Similarly as in case of elliptic flow, we can study correlations of the initial state eccentricity along the space-time rapidity. Analogous to the Eq. \eqref{eq:rn-long} we define the factorization ratio
\begin{equation}
r^\epsilon_n(\eta_s)=\frac{\left\langle \epsilon_n(-\eta_s)\epsilon_n(\etasref) \cos[n\left(\Psi_n(-\eta_s)-\Psi_n(\etasref)\right)]\right\rangle}{\left\langle \epsilon_n(\eta_s)\epsilon_n(\etasref) \cos[n\left(\Psi_n(\eta_s)-\Psi_n(\etasref)\right)]\right\rangle}.
\label{eq:rn-eps}
\end{equation}
For the  calculations we use the same intervals of $\eta_s$ as we did before with the pseudorapidity $\eta$.
Figure \ref{fig:eps2-decorrealtion} shows the factorization ratio of the initial state eccentricity calculated with both UrQMD and GLISSANDO initial states. From this figure one can see how the strong decorrelation, which is present in all previous results, originates from the initial state. 
Comparison with Fig.~\ref{fig:r2-27} shows that $r_2$ and $r_2^\varepsilon$ even almost agree quantitatively! However, note that the $r_2^\varepsilon$ is a coordinate-space characteristic of the initial state, whereas $r_2$ is a final-state momentum-space observable. On a qualitative level, such agreement is a result of hydrodynamic evolution which happens in between. That is similar to the established correspondence between the initial-state eccentricity $\epsilon_2$ and the final-state hadronic flow coefficient $v_2$, which is also well-explained in the hydrodynamic approach.

%%%%%%%%%%%%%%%%%%%%%

\subsection{Predictions for \AFTER}

\begin{figure}[tb]
    \centering
    \includegraphics[width=0.5\textwidth]{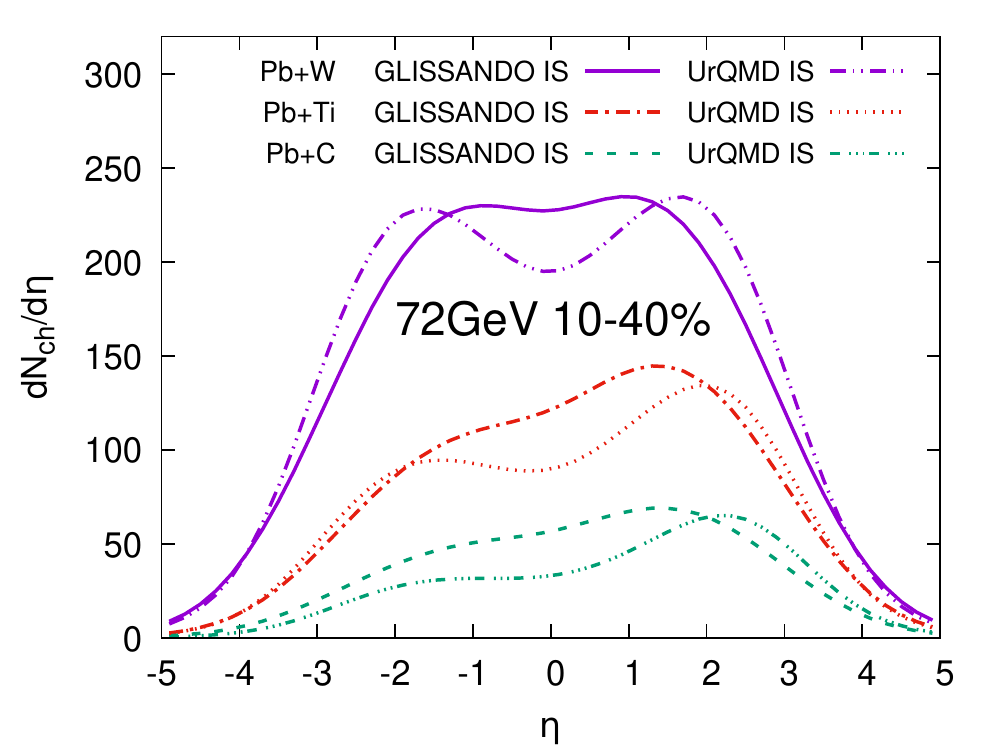}
    \caption{Prediction of pseudo-rapidity density of charged hadrons for 10-40\% Pb+W, Pb+Ti and Pb+C collisions at $\snn=72$~GeV from vHLLE+UrQMD simulations with UrQMD and GLISSANDO initial states.}
    \label{fig:dndeta-after}
\end{figure}

Finally, we calculated predictions for a proposed experiment \AFTER~\cite{Brodsky:2012vg,Massacrier:2015qba,Hadjidakis:2018ifr}. Since this is a fixed target experiment, it will have access to far backward rapidity region of the collision. Therefore, this experiment can be very useful tool to study the longitudinal structure of various observables.  We simulated collisions of Pb beam with Ti, W and C target at $\snn=72$~GeV. We start by calculating the basic observable: the pseudo-rapidity density of charged hadrons. In Fig.\ \ref{fig:dndeta-after} we show the pseudo-rapidity distributions for three colliding systems
%\del{setups} 
of \AFTER, calculated using both initial state models.

\begin{figure}[tb]
    \centering
    \includegraphics[width=0.5\textwidth]{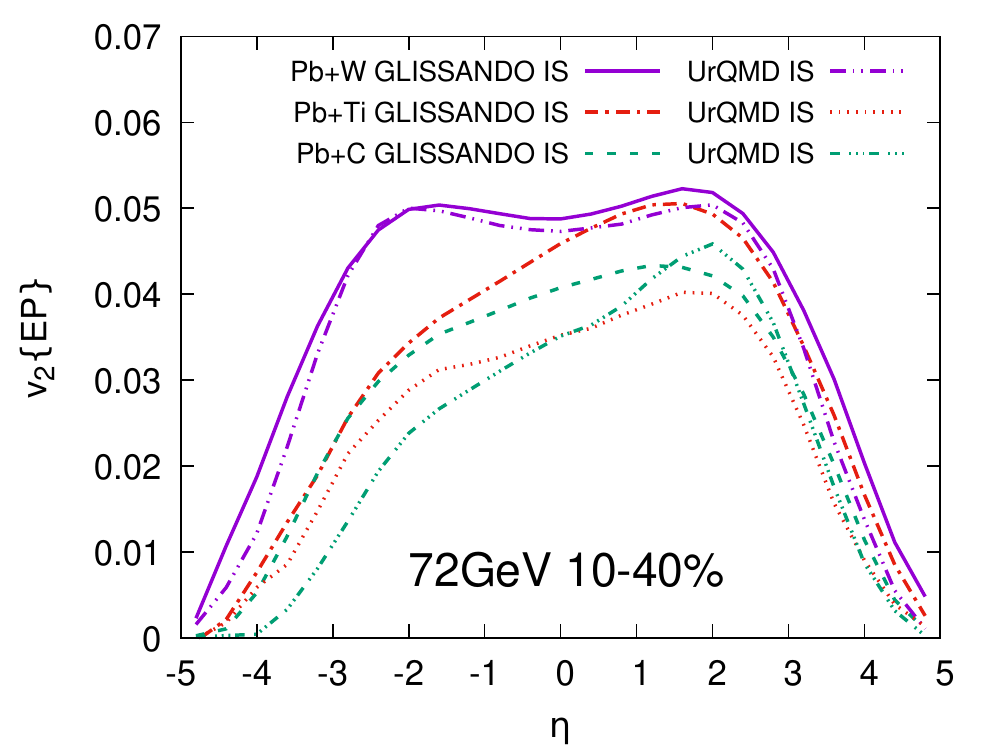}
    \caption{Prediction of elliptic flow as a function of pseudorapidity for 10-40\% Pb+W, Pb+Ti and Pb+C collisions at $\snn=72$~GeV from vHLLE+UrQMD simulations with UrQMD and GLISSANDO initial states.}
    \label{fig:v2-after}
\end{figure}

\begin{figure}[tb]
    \centering
    \includegraphics[width=0.5\textwidth]{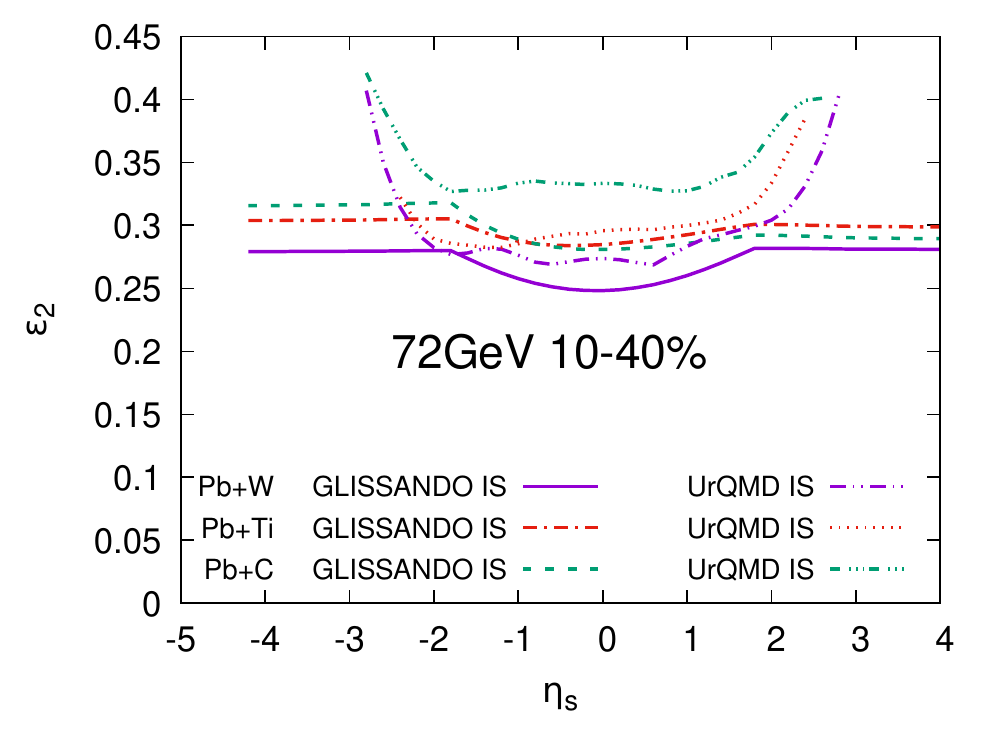}
    \caption{Initial state eccentricities as a functions of space-time rapidity for 10-40\% Pb+W, Pb+Ti and Pb+C collisions at $\snn=72$~GeV with UrQMD and GLISSANDO initial states.}
    \label{fig:eps2-after}
\end{figure}

\begin{figure}[tb]
    \centering
    \includegraphics[width=0.5\textwidth]{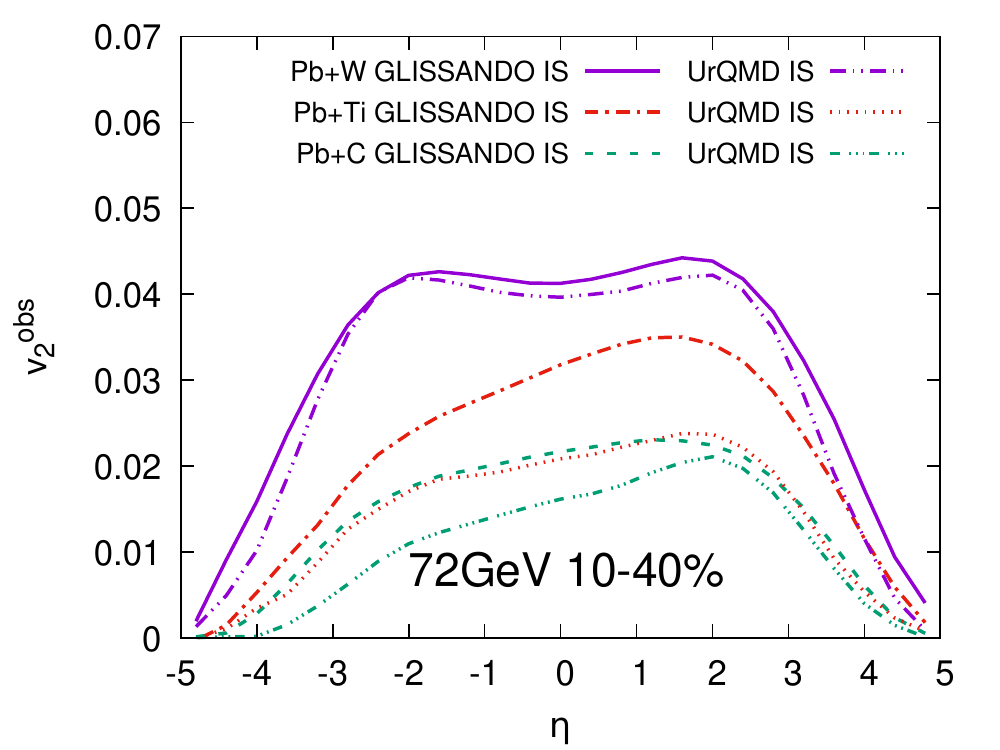}
    \caption{Same as Fig.~\ref{fig:v2-after} but without event-plane resolution correction.}
    \label{fig:v2-after-EP-Rcurly1}
\end{figure}

\begin{figure}[tb]
    \centering
    \includegraphics[width=0.5\textwidth]{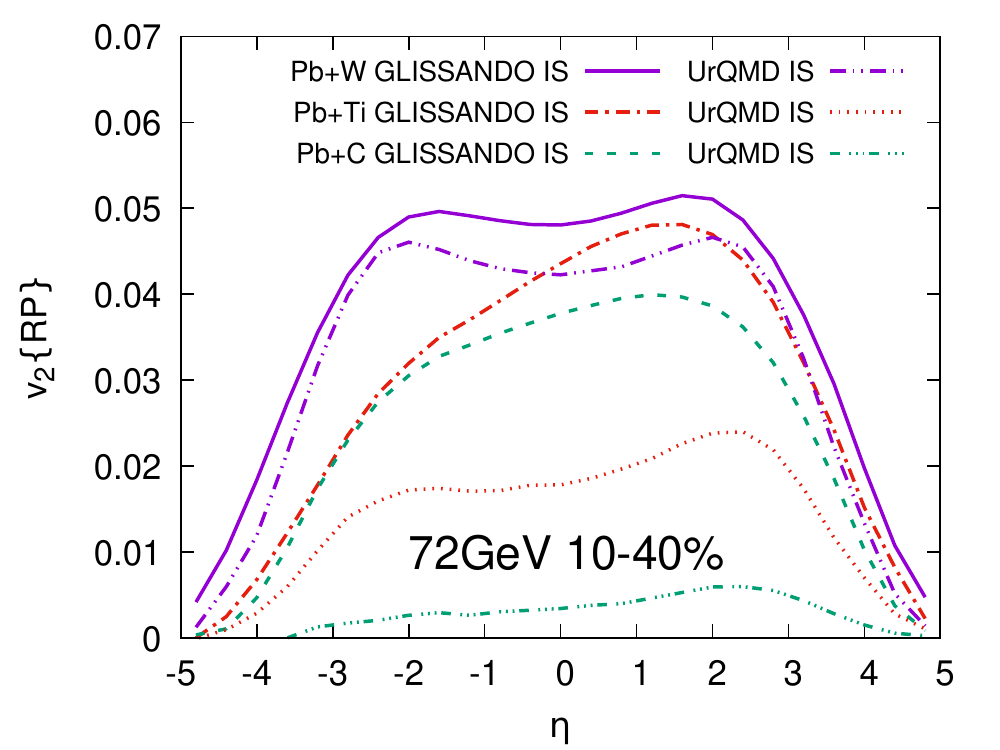}
    \caption{Same as Fig.~\ref{fig:v2-after} but computed with the reaction-plane method.}
    \label{fig:v2-after-RP}
\end{figure}

Next, we calculate the elliptic flow as a function of pseudorapidity, which is shown in Fig.\ \ref{fig:v2-after}. The biggest difference between the initial state models can be seen in Pb+Ti collisions, where on the Pb side UrQMD predicts much lower value of elliptic flow, even lower than in case of Pb+C collisions. 
To explain the peculiarity with the elliptic flow in Pb+Ti system, or in other words, a non-monotonic collision system size dependence, we should delve into the details of event-plane method used to compute  $v_2$. Pb-W, Pb-W and Pb-C systems have comparable initial-state eccentricities $\epsilon_2$ which in fact increase from Pb-W towards Pb-C system at the same centrality, see Fig.~\ref{fig:eps2-after}. However, smaller initial energy density for lighter systems prompts a shorter hydro phase, which develops a smaller final-state flow amplitude. 
On the other hand, note that in our faithful implementation of the event-plane method according to \cite{Holopainen:2010gz}, $v_2$ is obtained from the directly observed $v_2^{\rm obs}$ as 
\begin{equation}
v_2 = \frac{v_2^{\rm obs}}{\mathcal{R}}\,  ,
\end{equation}
where $\mathcal{R}$ is the event plane resolution. We have found that 
$\mathcal{R}$
decreases significantly from 0.84 for the heavy-heavy Pb+W down to 0.53 or 0.46 for the heavy-light Pb+C system, with GLISSANDO or UrQMD IS, respectively, as one can see in Table~\ref{tab:resolution-correction}. The latter is a result of the significant decrease in charged-hadron multiplicity from heavy-heavy to heavy-light system, see Fig.~\ref{fig:dndeta-after}, which makes the event-plane resolution less precise. Note that the system-size dependence of $\mathcal{R}$ is stronger in simulations with UrQMD IS as compared to the ones with GLISSANDO IS. As we plot the so-called $v_2^{\rm obs}$, which is $v_2\{\text{EP}\}$ without the event-plane resolution correction ($\mathcal{R}=1$) in Fig.~\ref{fig:v2-after-EP-Rcurly1}, we observe that the hierarchy $v_2^{\rm obs}(\text{Pb-W})>v_2^{\rm obs}(\text{Pb-Ti})>v_2^{\rm obs}(\text{Pb-C})$ is restored. However, the difference between GLISSANDO and UrQMD IS is still the largest for the Pb+Ti system. Furthermore, when we plot the reaction-plane $v_2$, see Fig.~\ref{fig:v2-after-RP}, we find that the difference between GLISSANDO and UrQMD IS grows from Pb+W via Pb+Ti to Pb+C system, where it is the largest. Another observation from Fig.~\ref{fig:v2-after-RP} is that the system-size dependence of $v_2\{\text{RP}\}$ is much weaker with GLISSANDO IS, as compared to UrQMD IS. The latter is again motivated by the structure of the initial state: a relatively small number of initial-state hadrons in Pb+C scenario, together with the relatively strong decorrelation of transverse energy density profiles at different rapidities, lead to a weak alignment of the flow vector with the original geometry of the collision, which causes a particularly small $v_2\{\text{RP}\}$ for Pb+C system. As such, the latter behaves like a ``small system'' rather than a classical heavy-ion system, from the point of view of elliptic flow. Therefore, the apparent non-monotonic system size dependence for the difference in $v_2\{{\rm EP}\}$ between UrQMD and GLISSANDO IS seen in Fig.~\ref{fig:v2-after} stems from an interplay between the different resolution corrections for different IS and colliding systems, general decrease of the flow signal from heavy-heavy to heavy-light system and the increase of the fluctuation-driven $v_2$ for the heavy-light system. The event-plane $v_2$ generally demonstrates  that the method becomes less reliable for the heavy-light collisions at the \AFTER\ energy, therefore more advanced methods such as $n-$particle cumulants should be used instead, which in turn would require much larger event statistics.

\begin{table}[tb]
 \centering
 \begin{tabular}{ c | c | c | c }
 %\begin{tabular}{ p{2.1cm} || p{1.1cm} | p{1.1cm} | p{1.1cm} }
 IS / system &  Pb-W  &  Pb-Ti  &  Pb-C \\ \hline \hline
 GLISSANDO &  0.846  &  0.693  &  0.532 \\
 UrQMD     &  0.838  &  0.591  &  0.460 \\
 \end{tabular}
    \caption{Resolution correction factors $\mathcal{R}$ in the event-plane $v_2$ for different collision systems at \AFTER energy.}
    \label{tab:resolution-correction}
\end{table}

\begin{figure}[tb]
\centering
\includegraphics[width=0.5\textwidth]{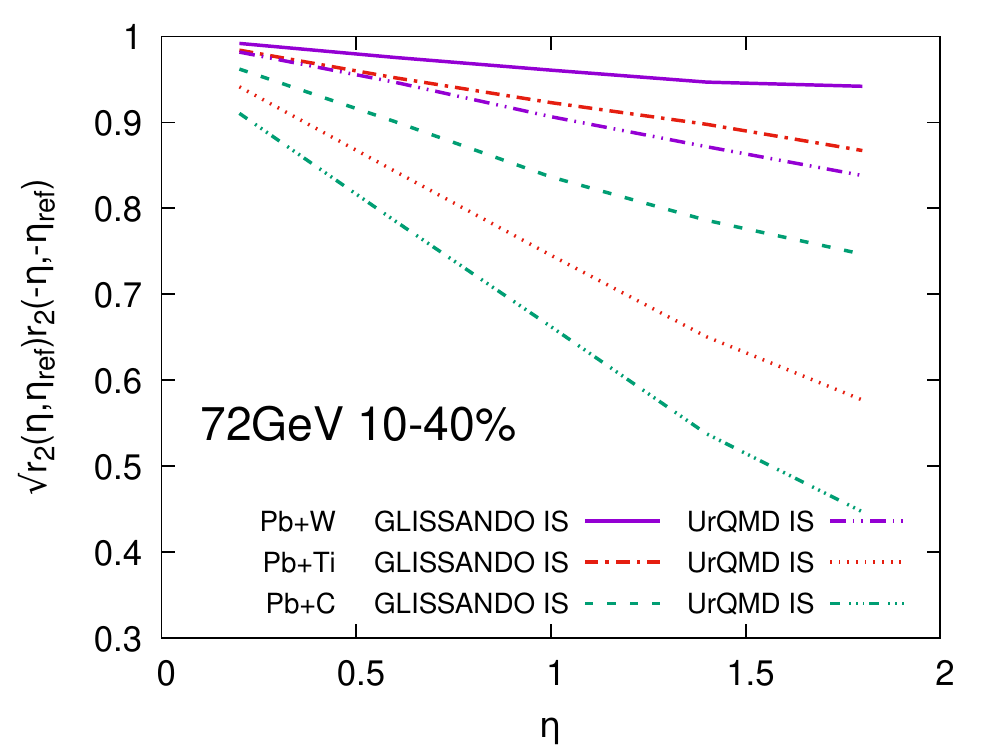}
\caption{Prediction of the symmetric factorization ratio $\sqrt{r_2(\eta, \etaref)r_2(-\eta,-\etaref)}$ as a function of pseudorapidity for 10-40\% Pb+W, Pb+Ti and Pb+C collisions at $\snn=72$~GeV from vHLLE+UrQMD simulations with UrQMD and GLISSANDO initial states.}
\label{fig:r2-after}
\end{figure}

Lastly, we calculate the prediction of the flow decorrelation at \AFTER. However, these collisions are asymmetric. In order to suppress the influence of the broken symmetry on the decorrelation measure, instead of using  Eq.~\eqref{eq:rn} we follow the work by CMS collaboration \cite{Khachatryan:2015oea} and take a product of $r_n(\eta)$ and $r_n(-\eta)$:
\begin{align}
&\sqrt{r_n(\eta, \etaref)r_n(-\eta,-\etaref)}= \nonumber \\ 
&\sqrt{\frac{\left\langle q_n(-\eta)q_n^\ast(\etaref) \right\rangle}{\left\langle q_n(\eta)q_n^\ast(\etaref) \right\rangle}\frac{\left\langle q_n(\eta)q_n^\ast(-\etaref) \right\rangle}{\left\langle q_n(-\eta)q_n^\ast(-\etaref) \right\rangle}}.
\label{eq:rn-symmetric}
\end{align}
Thanks to this we can study flow decorrelation in asymmetric collisions. For the calculation at \AFTER{}  energy we first use a reference pseudorapidity interval $2.1<\etaref<5.1$ and transverse momentum cut $0.4<p_T<4$~GeV/$c$. The resulting symmetric factorization ratio is showed in Fig.\ \ref{fig:r2-after}. In this figure, consistently with the results for $\snn=27$ and 200 GeV, we observe a significant difference between the two initial state models. Therefore, the flow decorrelation may be used as a tool to discriminate the models of the initial state. Also, both models agree that the decorrelation becomes stronger in smaller collision systems.

\begin{figure}[tb]
\centering
\includegraphics[width=0.5\textwidth]{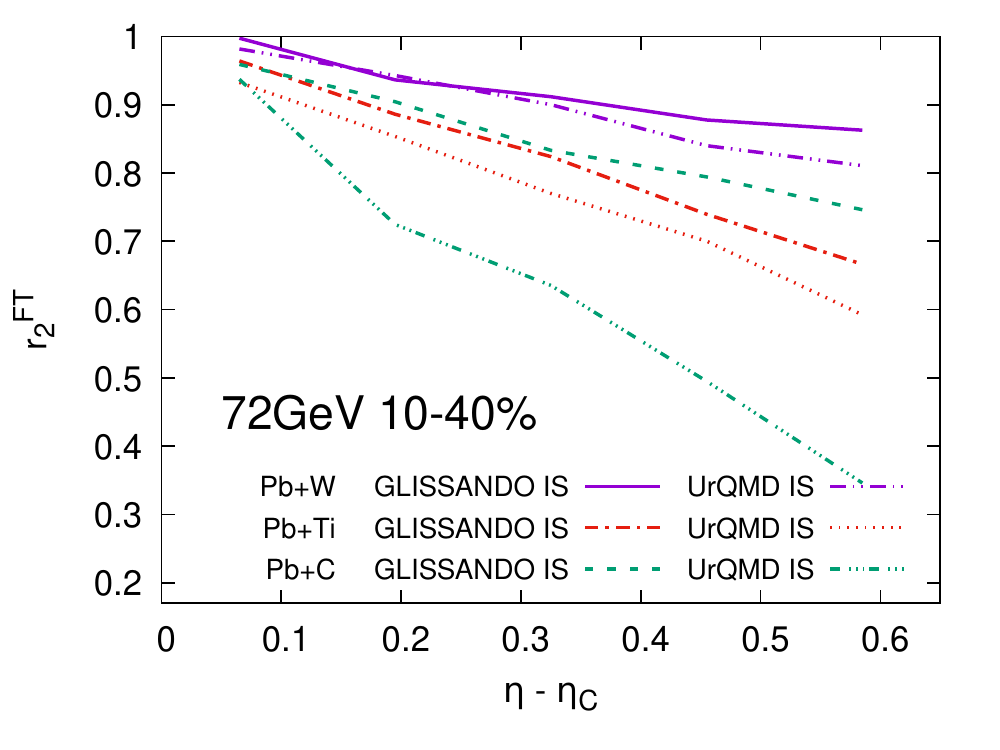}
\caption{Prediction of the factorization ratio $r^{\rm FT}_2$ as a function of $\eta-\eta_C$ for 10-40\% Pb+W, Pb+Ti and Pb+C collisions at $\snn=72$~GeV from vHLLE+UrQMD simulations with UrQMD and GLISSANDO initial states.}
\label{fig:r2-after2}
\end{figure}

Unfortunately, as a fixed-target experiment \AFTER\ does not provide access to the full mid- and forward rapidity region in the center-of-mass frame. Thus, the decorrelation cannot be measured using the classical definition, established for collider experiments for asymmetric systems. Therefore, we need to amend the definition of the factorization ratio for the fixed-target setup. The \AFTER\ experiment is projected to have two acceptance windows: $-1.0<\etaref<-0.5$, which can be used as reference range, and $-2.9<\eta<-1.6$, which can be used for the measurement itself. As such, we calculate the decorrelation around the center of the pseudo-rapidity bin $\eta_C=-2.25$ as follows:
\begin{equation}
r^{\rm FT}_n(\eta-\eta_C)= 
\frac{\langle q_n(-\eta+2 \eta_C)q_n^*(\etaref) \rangle}%
{\langle q_n(\eta)q_n^*(\etaref) \rangle}\, .
\label{eq:rn_after}
\end{equation}
Figure \ref{fig:r2-after2} shows the factorization ratio calculated using Eq.\ \eqref{eq:rn_after}. These results can be reproduced in experimental conditions at \AFTER. 
%\del{Also, we generally observe a stronger decorrelation around $\eta_C$, as compared to the one around midrapidity. {\bf -- apart from} Pb+C UrQMD case, it looks the same actually.}
Here, simulations with UrQMD IS show a steady decrease of flow correlation from Pb+W towards Pb+C system. However, simulations with GLISSANDO IS result in $r_2^{\rm FT}$ which is slightly larger for Pb+C system as compared to Pb+Ti. 

\begin{figure}[tb]
\centering
\includegraphics[width=0.5\textwidth]{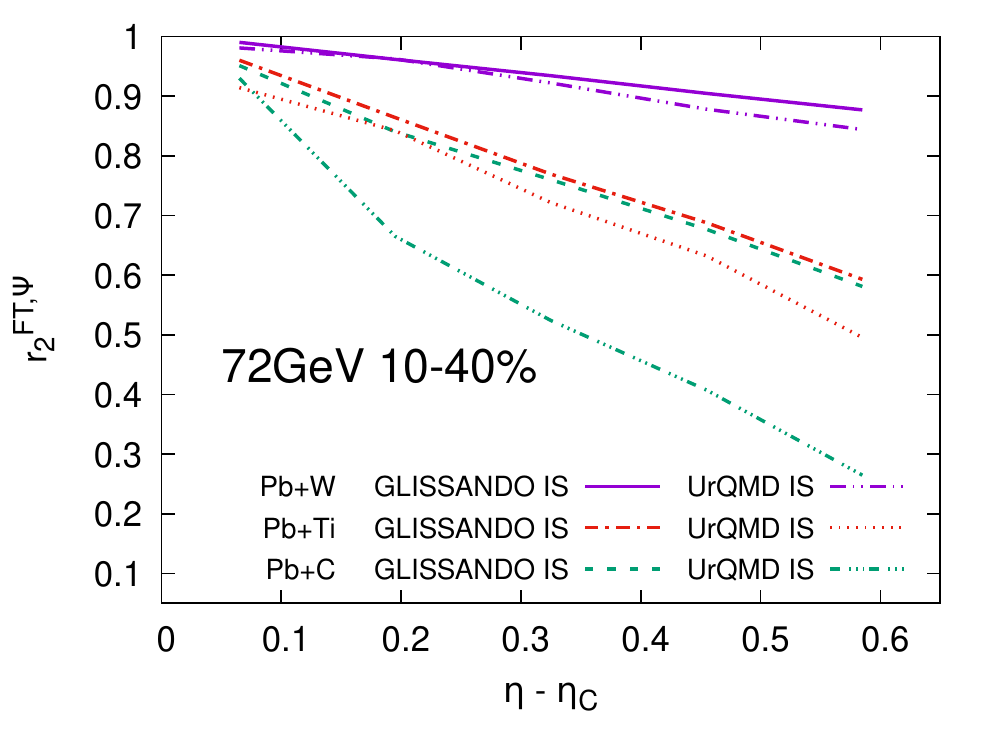}
\caption{Flow angle decorrelation $r^{\rm FT,\Psi}_2$ as a function of $\eta-\eta_C$ for 10-40\% Pb+W, Pb+Ti and Pb+C collisions at $\snn=72$~GeV from vHLLE+UrQMD simulations with UrQMD and GLISSANDO initial states.}
\label{fig:r2psi-after2}
\end{figure}

In order to explain such irregularity, we can---in analogy to eqs.~(\ref{eq:rn-v}) and (\ref{eq:rn-psi})---introduce the 
decorrelation in the magnitude and the flow angle around $\eta_C$
\begin{eqnarray}
r^{{\rm FT},v}_n(\eta-\eta_C) & = & 
\frac{\langle v_n(-\eta+2 \eta_C)v_n^*(\etaref) \rangle}%
{\langle v_n(\eta)v_n^*(\etaref) \rangle}
\\
r^{{\rm FT},\Psi}_n(\eta-\eta_C) &= &
\frac{\left\langle \cos[n\left(\Psi_n(-\eta+2\eta_C)-\Psi_n(\etaref)\right)]\right\rangle}{\left\langle  \cos[n\left(\Psi_n(\eta)-\Psi_n(\etaref)\right)]\right\rangle}\,  .
\end{eqnarray}
The flow angle decorrelation is shown in Fig.~\ref{fig:r2psi-after2}.
%\del{we compute the flow angle decorrelation $r_2^{\rm FT,\Psi}$ around $\eta_C$, computed for the same scenarios, and show it on Fig.~\ref{fig:r2psi-after2}. }
One can see from the figure that $r_2^{\rm FT,\Psi}$ retains monotonic system-size dependence for both IS scenarios, however it seems to change very little between Pb+Ti and Pb+C systems. On the other hand, from Fig.~\ref{fig:v2-after-EP-Rcurly1} one can see that within $-2.9<\eta<-1.6$, the magnitude of the $q_2$ vector has a steeper pseudorapidity dependence for Pb+Ti system as compared to Pb+C. 
Therefore, one can expect that the decorrelation of the flow magnitude $r_2^{\rm FT,v}$ will drop with $|\eta-\eta_C|$ faster for Pb+Ti than for Pb+C system. Hence, the non-monotonic system-size dependence of $r_2^{\rm FT}$ stems from different slopes of the pseudorapidity dependence of $v_2$ for Pb+Ti and Pb+C systems. 

%%%%%%%%%%%%%%%%%%%%%%%%%%%%%%%%%%%%%%%

\section{Conclusions}
We have presented the rapidity-dependent elliptic flow and flow decorrelation in Au-Au collisions at $\snn=27$~and 200 GeV at RHIC, as well as in Pb-Ti, Pb-W and Pb-C collisions at $\snn=72$ GeV in the \AFTER{} experiment, computed in viscous hydrodynamic + cascade model with UrQMD and 3D GLISSANDO initial states. We found that, in all cases, UrQMD IS results in much stronger final-state elliptic flow decorrelation as compared to 3D GLISSANDO IS. At $\snn=27$~GeV, the flow decorrelation with UrQMD IS is stronger than the preliminary data from STAR \cite{Nie:2020trj}, whereas 3D GLISSANDO IS agrees with the data within error bars. The difference is rooted in a much stronger decorrelation of initial-state eccentricity as a function of space-time rapidity in UrQMD IS, as compared to 3D GLISSANDO IS. Similarly to the previous findings, we observe that the dominant effect in the flow decorrelation is decorrelation of flow angle, or $r_2^\Psi$. As an extension over the existing decorrelation studies, we demonstrate the effect of final-state hadronic rescatterings, and find that the rescatterings suppress flow decorrelation: $r_2$ comes closer to unity.

\begin{acknowledgments}
JC, IK, and BT acknowledge support by the project Centre of Advanced Applied Sciences, No.~CZ.02.1.01/0.0/0.0/16-019/0000778,  co-financed by the European Union. JC and BAT acknowledge support from from The Czech Science Foundation, grant number: GJ20-16256Y. IK acknwowledges support by the Ministry of Education, Youth and Sports of the Czech Republic under grant ``International Mobility of Researchers – MSCA IF IV at CTU in Prague'' No.\ CZ.02.2.69/0.0/0.0/20\_079/0017983. BT acknowledges support from VEGA 1/0348/18. Computational resources were supplied by the project ``e-Infrastruktura CZ'' (e-INFRA LM2018140) provided within the program Projects of Large Research, Development and Innovations Infrastructures.
\end{acknowledgments}

\bibliographystyle{h-physrev.bst}
\bibliography{main}

\end{document}